\documentclass[preprint,showpacs,preprintnumbers,amsmath,amssymb]{revtex4}
\usepackage{epsfig}

\newcommand{\logtwo}{l_2}
\newcommand{\logtri}{l_3}
\newcommand{\logR}{l_R}

\newcommand{\bea}{\begin{eqnarray}}
\newcommand{\eea}{\end{eqnarray}}

\newcommand{\re}{\mathop{\mathrm{Re}}\nolimits}
\newcommand{\Li}{\mathop{\mathrm{Li}}\nolimits}
\newcommand{\Si}{\mathop{\mathrm{S}}\nolimits}

\begin{document}

\preprint{DESY~09--136\hspace{12cm} ISSN 0418--9833}
\preprint{September 2009\hspace{14.5cm}}

\boldmath
\title{Orthopositronium lifetime at ${\mathcal O}(\alpha)$ and
${\mathcal O}(\alpha^3\ln\alpha)$ in closed form}
\unboldmath

\author{B.A.~Kniehl, A.V.~Kotikov\footnote{%
On leave of absence from Bogoliubov Laboratory for Theoretical Physics, JINR,
141980 Dubna (Moscow region), Russia.}
and O.L.~Veretin}
\affiliation{{II.} Institut f\"ur Theoretische Physik, Universit\"at Hamburg,
22761 Hamburg, Germany}

\date{\today}

\begin{abstract}
Recently, the ${\mathcal O}(\alpha)$ and ${\mathcal O}(\alpha^3\ln\alpha)$
radiative corrections to the orthopositronium lifetime have been presented in
closed analytical form, in terms of basic irrational numbers that can be
evaluated numerically to arbitrary precision
[Phys.\ Rev.\ Lett.\  {\bf 101}, 193401 (2008)].
Here, we present the details of this calculation and reveal the nature of
these new constants.
We also list explicit transformation formulas for generalized polylogarithms
of weight four, which may be useful for other applications.

\end{abstract}

\pacs{12.20.Ds, 31.30.J-, 36.10.Dr}
%%\keywords{Suggested keywords}
\maketitle

\section{Introduction}

Positronium (Ps), the electron-positron bound state, was discovered 
experimentally in 1951 \cite{Deutsch:1951zza}.
Since that time a lot of attention has been paid to the determination of its
properties, including lifetime, decay modes, and spectroscopy.
The experimental and theoretical accuracies achieved by now being quite high,
there is little doubt that quantum electrodynamics (QED) is the only
interaction in this system.
In fact, thanks to the smallness of the electron mass $m$ relative to
typical hadronic mass scales, its theoretical description is not plagued by
strong-interaction uncertainties and its properties, such as decay widths and
energy levels, can be calculated perturbatively in non-relativistic QED
(NRQED) \cite{Caswell:1985ui}, as expansions in Sommerfeld's fine-structure
constant $\alpha$, with very high precision.

Ps comes in two ground states, $^1S_0$ parapositronium ($p$-Ps) and $^3S_1$
orthopositronium ($o$-Ps), which decay to two and three photons, respectively.
Here we are concerned with the lifetime of $o$-Ps, which has been the subject
of a vast number of experimental and theoretical investigations.
Its first measurement \cite{Deutsch:1951zz} was performed later in the year
1951 and agreed well with its lowest-order (LO) prediction of 1949
\cite{Ore:1949te}.
Its first precision measurement \cite{BH}, of 1968, had to wait 9 years for
the first correct one-loop calculation \cite{Caswell:1976nx}, which came two
decades after the analogous calculation for $p$-Ps \cite{Harris:1957zz} being
considerably simpler owing to the two-body final state.
In the year 1987, the Ann Arbor group \cite{Westbrook:1987zz} published a
measurement that exceeded the best theoretical prediction available then by
more than 8 experimental standard deviations.
This so-called $o$-Ps lifetime puzzle triggered an avalanche of both
experimental and theoretical activities, which eventually resulted in what now
appears to be the resolution of this puzzle.
In fact, the 2003 measurements at Ann Arbor \cite{Vallery:2003iz} and Tokyo
\cite{Jinnouchi:2003hr},
\begin{eqnarray}
\Gamma(\mbox{Ann Arbor}) &=&
7.0404(10~\mbox{stat.})(8~\mbox{syst.})~\mu s^{-1},
\nonumber\\
\Gamma(\mbox{Tokyo}) &=&
7.0396(12~\mbox{stat.}) (11~\mbox{syst.})~\mu s^{-1},  
\end{eqnarray}
agree mutually and with the present theoretical prediction,
\begin{equation}
\Gamma(\mbox{theory}) = 7.039979(11)~\mu s^{-1}.
\end{equation}
The latter is evaluated from
\begin{equation}
\Gamma(\mbox{theory})=\Gamma_0\left[1 + A \frac{\alpha}{\pi}
+\frac{\alpha^2}{3} \ln\alpha 
+B \left(\frac{\alpha}{\pi}\right)^2
- \frac{3\alpha^3}{2\pi} \ln^2 \alpha 
  + C \frac{\alpha^3}{\pi} \ln \alpha   \right],
\label{Gamma}
\end{equation}
where \cite{Ore:1949te}
\begin{equation}
\Gamma_0 = \frac{2}{9}(\pi^2-9)\frac{m\alpha^6}{\pi}
\end{equation}
is the LO result.
The leading logarithmically enhanced ${\mathcal O}(\alpha^2\ln\alpha)$ and
${\mathcal O}(\alpha^3\ln^2\alpha)$ terms were found in
Refs.~\cite{Caswell:1978vz,Khriplovich:1990eh} and Ref.~\cite{Kar},
respectively.
The coefficients $A=-10.286606(10)$
\cite{Caswell:1976nx,Caswell:1978vz,Stroscio:1974zz,Adkins:2000fg,%
Adkins:2005eg},
$B=45.06(26)$ \cite{Adkins:2000fg},
and $C=-5.51702455(23)$ \cite{Kniehl:2000dh} were evaluated numerically in a
series of papers.
Comprehensive reviews of the experimental and theoretical status of Ps may be
found in Refs.~\cite{AFS,Rubbia:2004ix}.

We note in passing that high-precision tests make Ps also a useful probe of
new physics beyond he standard model.
At present, there is strong interest in models with extra dimensions
\cite{Akama:1982jy}, which may provide a solution of the gauge hierarchy
problem \cite{ArkaniHamed:1998rs} (see Ref.~\cite{Rubakov:2001kp} for a
review).
Some time ago, a peculiar feature of matter in brane world was observed in
Ref.~\cite{Dubovsky:2000am}, where it was shown that massive particles
initially located on our brane may leave the brane and disappear into extra
dimensions.
The experimental signature of this effect is the disappearance of a
particle from our world, {\it i.e.}\ its invisible decay.
The case of the electromagnetic field propagating in the Randall--Sundrum
type of metric in the presence of extra compact dimensions 
\cite{Oda:2000zc,Dubovsky:2000av} was considered in
Ref.~\cite{Dubovsky:2000av}, where it was shown that the transition rate of a
virtual photon into extra dimensions is non-zero.
This effect could result in the disappearance of a neutral system.
In the case of $o$-Ps, such estimations for the invisible decay branching
fraction $B(o$-Ps${}\to{}${\it invisible})
\cite{Gninenko:2003nx,Rubbia:2004ix} range just one order of magnitude below
the presently best experimental upper bound of $4.3\times10^{-7}$ at 90\%
confidence level established by Badertscher {\it et al.}\
\cite{Badertscher:2006fm}.
Thus, this decay is of great interest for the possible observation of effects
due to extra dimensions.

In order to reduce the theoretical uncertainty in the $o$-Ps total decay width
$\Gamma$(theory), it is indispensable to increase the precision in the
coefficients $A$, $B$, and $C$ in Eq.~(\ref{Gamma}).
This is most efficiently done by avoiding numerical integrations altogether,
{\it i.e.}\ by establishing the analytic forms of these coefficients.
The case of $B$ is beyond the scope of presently available technology, since
it involves two-loop five-point functions to be integrated over the
three-particle phase space.
In the following, we thus concentrate on $A$ and $C$.
The quest for an analytic expression for $A$ has a long history.
About 25 years ago, some of the simpler contributions to $A$, due to
self-energy and outer and inner vertex corrections, were obtained analytically
\cite{Stroscio:1982wj}, but further progress then soon came to a grinding halt.
In our recent Letter \cite{Kniehl:2008ia}, this task was completed for $A$ as
a whole.
The purpose of the present paper is to explain the most important technical
details of this calculation and to collect mathematical identities that may be
useful for similar calculations.

An analytic expression for $C$ is then simply obtained from that for $A$
through the relationship \cite{Kniehl:2000dh}
\begin{equation}
C = \frac{A}{3}- \frac{229}{30} + 8 \ln2,
\label{Cres}
\end{equation}
which may be understood qualitatively by observing that the
${\mathcal O}(\alpha^3\ln\alpha)$ correction in Eq.~(\ref{Gamma}) receives a
contribution from the interference of the relativistic ${\mathcal O}(\alpha)$
term from the hard scale with non-relativistic
${\mathcal O}(\alpha^2\ln\alpha)$ terms from softer scales.

The structure of this paper is as follows.
Section~\ref{sec:app} contains the well-known integral representation of the
$o$-Ps total decay width as given in Ref.~\cite{Adkins:2005eg}.
In Sec.~\ref{sec:use}, we show how to transform the contributing integrals to
forms appropriate for analytic evaluation, which is carried out for the most
complicated integrals, which are plagued by singularities, in
Sec.~\ref{sec:eval}.
More examples are studied in Sec.~\ref{sec:some}.
The final results for the coefficients $A$ and $C$ are presented in
Sec.~\ref{sec:res}.
Section~\ref{sec:con} contains a summary.
In Appendix~\ref{sec:a}, we present the analytic results for all parts of the
integral representation given in Sec.~\ref{sec:app}.
Appendix~\ref{sec:b} contains useful representations of the $\psi$ function
and the expansion of the $\Gamma$ function about half-integer-valued
arguments.
In Appendix~\ref{sec:c}, transformation formulas for generalized
polylogarithms of weight four with different arguments are collected.

\section{Definitions and notations}
\label{sec:app}

\begin{figure}[ht]
\begin{center}
\includegraphics[width=0.45\textwidth]{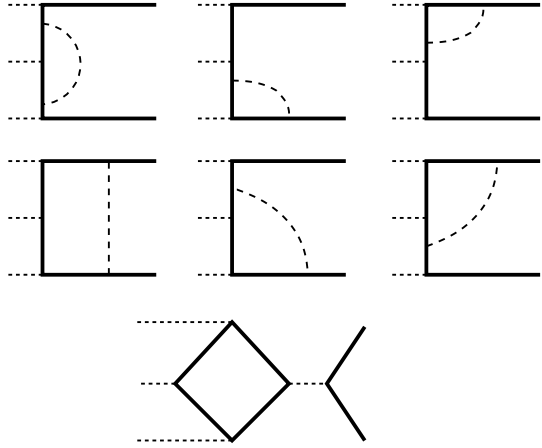}
\end{center}
\caption{\label{fig:dia}%
Feynman diagrams contributing to the total decay width of $o$-Ps at
${\mathcal O}(\alpha)$.
Self-energy diagrams are not shown.
Dashed and solid lines represent photons and electrons, respectively.}
\end{figure}
The ${\mathcal O}(\alpha)$ contribution in Eq.~(\ref{Gamma}),
$\Gamma_1=\Gamma_0A\alpha/\pi$, is due to the Feynman diagrams where a virtual
photon is attached in all possible ways to the tree-level diagrams, with three
real photons linked to an open electron line, and the electron box diagrams
with an $e^+e^-$ annihilation vertex connected to one of the photons being
virtual (see Fig.~\ref{fig:dia}).
Taking the interference with the tree-level diagrams, imposing $e^+e^-$
threshold kinematics, and performing the loop and angular integrations, one
obtains the two-dimensional integral representation \cite{Adkins:2005eg} 
\begin{equation}
\Gamma_1 = \frac{m\alpha^7}{36\pi^2}
\int\limits^1_0\frac{dx_1}{x_1}\,\frac{dx_2}{x_2} \,
\frac{dx_3}{x_3}\delta(2-x_1-x_2-x_3)
[F(x_1,x_3) + {\mathrm{perm.}}],
\label{eq:org}
\end{equation}
where $x_i$, with $0\le x_i\le 1$, is the energy of photon $i$ in the $o$-Ps
rest frame normalized to its maximum value, the delta function ensures energy
conservation, and ''perm.'' stands for the other five permutations of
$x_1,x_2,x_3$.
The function $F(x_1,x_3)$ is given by
\begin{equation}
F(x_1,x_3) = g_0(x_1,x_3) + \sum_{i=1}^5 g_i(x_1,x_3) h_i(x_1)
+ \sum_{i=6}^7 g_i(x_1,x_3) h_i(x_1,x_3),
\label{a.2}
\end{equation}
where $g_i(x_1,x_3)$ are ratios of polynomials, which are listed 
in Eqs.~(A5a)--(A5h) of Ref.~\cite{Adkins:2005eg}, and
\begin{eqnarray}
h_1(x_1) &=& \ln (2x_1),\qquad
h_2(x_1) = \sqrt{\frac{x_1}{\overline{x}_1}} \, \theta_1,\qquad
h_3(x_1) = \frac{1}{2x_1} [\zeta(2)-\Li_2(1-2x_1)],
\nonumber\\
h_4(x_1) &=& \frac{1}{4x_1} [3\zeta(2)-2\theta_1^2],\qquad
h_5(x_1) = \frac{1}{2\overline{x}_1} \theta_1^2,
\\
h_6(x_1,x_3) &=&
\frac{1}{\sqrt{x_1\overline{x}_1x_3\overline{x}_3}} \,
\left[\Li_2\left(r^+_A,\overline{\theta}_1\right)
-\Li_2\left(r^-_A,\overline{\theta}_1\right)\right],
\label{h6}\\
h_7(x_1,x_3) &=& \frac{1}{2\sqrt{x_1\overline{x}_1x_3\overline{x}_3}} \,
\left[ 2\Li_2(r^+_B,\theta_1) - 2\Li_2(r^-_B,\theta_1) - \Li_2(r^+_C,0)
+ \Li_2(r^-_C,0)\right],
\label{h7}
\end{eqnarray}
where $\overline{x}_i=1-x_i$ and
\begin{eqnarray}
\theta_1 &=& \arctan\sqrt{\frac{\overline{x}_1}{x_1}},\qquad
\overline{\theta}_1 = \arctan\sqrt{\frac{x_1}{\overline{x}_1}},\qquad
p_A = \sqrt{\frac{x_1\overline{x}_3}{\overline{x}_1x_3}},\qquad
p_B = \sqrt{\frac{\overline{x}_1\overline{x}_3}{x_1x_3}},
\nonumber \\
r^{\pm}_A &=& 
\sqrt{\overline{x}_1} \, (1 \pm p_A),\qquad
r^{\pm}_B = \sqrt{x_1} \, (1 \pm p_B),\qquad
r^{\pm}_C = \frac{r^{\pm}_B}{\sqrt{x_1}}.
 \label{1.3}
\end{eqnarray}
Here, $\zeta(2)=\pi^2/6$ and
\begin{equation}
\Li_2(r,\theta) =
-\frac{1}{2}\int\limits^1_0\frac{dt}{t}
\ln(1-2 r t\cos\theta+r^2 t^2)
 \label{1.4}
\end{equation}
is the real part of the dilogarithm [see line below Eq.~(\ref{eq:s})] of
complex argument $z=r{\mathrm e}^{{\mathrm i}\theta}$ \cite{Lewin}.
Since we are dealing here with a single-scale problem, Eq.~(\ref{eq:org})
yields just a number.

Although Bose symmetry is manifest in Eq.~(\ref{eq:org}), its evaluation is
complicated by the fact that, for a given order of integration, individual
permutations yield divergent integrals, which have to cancel in their
combination.
In order to avoid such a proliferation of terms, we introduce an infinitesimal
regularization parameter $\delta$ in such a way that the symmetry under
$x_i\leftrightarrow x_j$ for any pair $i\ne j$ is retained.
In this way, Eq.~(\ref{eq:org}) collapses to
\begin{equation}
\Gamma_1 = \frac{m\alpha^7}{6\pi^2}
\int\limits^{1-\delta}_{2\delta} dx_{1}
\int\limits^{1-\delta}_{1-x_1+\delta}
\frac{dx_2}{x_1x_2x_3}F(x_1,x_3),
\label{a.1}
\end{equation}
where $x_3=2-x_1-x_2$.
Note that we may now exploit the freedom to choose any pair of variables $x_i$
and $x_j$ $(i\neq j)$ as the arguments of $F$ and as the integration
variables.

\section{Integral representations of dilogarithmic functions}
\label{sec:use}

Obviously, the functions $h_6(x_1,x_3)$ and $h_7(x_1,x_3)$ in Eqs.~(\ref{h6})
and (\ref{h7}), respectively, give the most complicated contributions
to $\Gamma_1$.
In order to perform integrations involving these terms, it is useful to apply 
the integral representation of
Eq.~(\ref{1.4}) to $\Li_2(r^{\pm}_A,\overline{\theta}_1)$,
$\Li_2(r^{\pm}_B,\theta_1)$, and $\Li_2(r^{\pm}_C,0)$.
Let us first consider $\Li_2(r^+_B,\theta_1)$.
We see from Eq.~(\ref{1.3}) that $\cos\theta_1=\sqrt{x_1}$ and thus
\begin{equation}
\Li_2(r^+_B, \theta_1) =
-\frac{1}{2} \int\limits^{1+p_B}_0 \frac{dt_1}{t_1} 
\ln[1- x_1 t_1(2-t_1)],
 \label{1.5}
\end{equation}
where $t_1=(1+p_B)t$.
Then, the term $D_1=\Li_2(r^+_B, \theta_1)-\Li_2(r^-_B, \theta_1)$ on the
r.h.s.\ of Eq.~(\ref{h7}), after the change $t_2=t_1-1$, can be rewritten as
\begin{equation}
D_1 =
-\frac{1}{2} \int\limits^{p_B}_{-p_B} \frac{dt_2}{1+t_2} 
\ln[1- x_1 (1-t_2^2)].
\label{1.6}
\end{equation}
Finally, substituting $t_2=p_B \sqrt{t}$, we obtain
\begin{equation}
D_1 =  -\frac{1}{2} 
\sqrt{x_1\overline{x}_1x_3\overline{x}_3} \int\limits^{1}_{0} 
\frac{dt}{\sqrt{t}(x_1x_3 - \overline{x}_1\overline{x}_3 t)} 
[\ln \overline{x}_1 -\ln x_3 + \ln(x_3+ \overline{x}_3 t)].
 \label{1.7}
\end{equation}
The residual term on the r.h.s.\ of Eq.~(\ref{h7}),
$D_2=\Li_2(r^+_C, 0)-\Li_2(r^-_C, 0)$, can be transformed in the same way
yielding
\begin{equation}
D_2 = -\frac{1}{2}  
\sqrt{x_1\overline{x}_1x_3\overline{x}_3} \int\limits^{1}_{0} 
\frac{dt}{\sqrt{t}(x_1x_3 - \overline{x}_1\overline{x}_3 t)}
[\ln(\overline{x}_1 \overline{x}_3) - \ln(x_1 x_3) + \ln t].
\label{1.8}
\end{equation}
We thus obtain the following integral representation for
$h_7(x_1,x_3)$ \footnote{%
Equation~(\ref{1.5a}) corrects a misprint in Eq.~(15) of
Ref.~\cite{Kniehl:2008ia}, which is, however, inconsequential because the
difference between these expressions cancels due to the
$x_1 \leftrightarrow x_3$ symmetry.}:
\begin{equation}
h_7(x_1,x_3) = -\frac{1}{4} \int\limits^1_0 \frac{dt}{\sqrt{t}
\left(x_1x_3-\overline{x}_1\overline{x}_3 t\right)}
\left[\ln\frac{\overline{x}_1x_1}{x_3\overline{x}_3}
+2 \ln(x_3+\overline{x}_3 t)-\ln t \right].
\label{1.5a}
\end{equation}
Exploiting the $x_1 \leftrightarrow x_3$ symmetry of the coefficient
$g_7(x_1,x_3)$ multiplying $h_7(x_1,x_3)$, Eq.~(\ref{1.5a}) can be effectively
replaced by
\begin{equation}
\tilde{h}_7(x_1,x_3) = -\frac{1}{4} \int\limits^1_0 \frac{dt}{\sqrt{t}\left(x_1x_3-
\overline{x}_1\overline{x}_3 t\right)}
[2\ln(x_3+\overline{x}_3 t) - \ln t].
  \label{1.6a}
\end{equation}
Next, this expression, multiplied by $g_7(x_1,x_3)$, is to be integrated
over $x_1$, $x_3$, and $t$.
Observing that the logarithmic terms in Eq.~(\ref{1.6a}) are independent of
$x_1$, we first integrate over $x_1$ (for a similar approach, see
Ref.~\cite{Kniehl:2005yc}).
In order to avoid the appearance of complicated functions in the intermediate
results, the integration over $t$ in Eq.~(\ref{1.6a}) is performed last.

Using the same technique, we obtain the following representation for the
function $h_6(x_1,x_3)$:
\begin{equation}
\tilde{h}_6(x_1,x_3) = -\frac{1}{2} \int\limits^1_0 
  \frac{dt}{\sqrt{t} ( \overline{x}_1x_3-
          x_1\overline{x}_3 t)} [\ln x_1 - \ln x_3 +
\ln(x_3+\overline{x}_3 t) ] ,
  \label{1.7a}
\end{equation}
in which the part proportional to $\ln x_1$ and the complementary one are
first integrated over $x_3$ and $x_1$, respectively.
The $t$ integration is again performed last. 

In Secs.~\ref{sec:eval} and \ref{sec:some}, we discuss in more details how
these integrations can be performed.

%The above procedure allows one to obtain the result for these integrals in
%terms of a infinite sum, which can be summed afterwards to some known
%constants.

\boldmath
\section{Evaluation of contributions with $h_6$ and $h_7$}
\unboldmath
\label{sec:eval}

We now discuss the evaluation of the most complicated integrals, namely those
involving the functions $h_6(x_1,x_3)$ and $h_7(x_1,x_3)$.
We denote the corresponding integrated expressions as $I_6$ and $I_7$,
respectively.
They are both singular for $\delta \to 0$, so that the regularization of
Eq.~(\ref{a.1}) is indispensable.

Let us first consider the contribution of the coefficient $g_7(x_1,x_3)$
without the function $h_7(x_1,x_3)$.
It can be decomposed into two parts, as 
\begin{equation}
\tilde g_7(x_1,x_3)=\frac{g_7(x_1,x_3)}{x_1x_3(2-x_1-x_3)} 
= \tilde{g}^{\rm{sing}}_7(x_1,x_3) + \tilde{g}^{\rm{reg}}_7(x_1,x_3),
  \label{1b.1}
\end{equation}
where
\begin{equation}
\tilde{g}^{\rm{sing}}_7(x_1,x_3) = \frac{3x_3(1-x_3)}{2-x_1-x_3}
  \label{1b.2}
\end{equation}
gives rise to the singularity upon integration over $x_1$ and $x_3$, while
\begin{eqnarray}
\tilde{g}^{\rm{reg}}_7(x_1,x_3) &=& 
   \frac{18}{x_3} - 3 + 9 x_3 + \left( \frac{2}{x_3} - 10 \right) x_1
   + \left( \frac{4}{2-x_3} - \frac{8}{x_3} + 10 + 2 x_3 \right) \frac{1}{x_1}
   \nonumber\\
  &&{}+
   \left( -\frac{52}{2-x_3} - \frac{12}{x_3} + 66 - 44 x_3 + 11 x_3^2 \right)
   \frac{1}{2-x_1-x_3}
\end{eqnarray}
remains finite, so that the limit $\delta\to0$ can be taken. 
A similar decomposition can be made also for $g_6(x_1,x_3)$.
Specifically, performing the integrations over $x_1$ and $x_3$ and taking the
limit $\delta\to0$, we have
\begin{eqnarray}
6 \int\limits^{1-\delta}_{2\delta} dx_{1} 
\int\limits^{1-\delta}_{1-x_1+\delta} 
dx_3 \, \tilde{g}^{\rm{sing}}_7(x_1,x_3) &=& 3 \ln\delta + \frac{5}{2} 
    + {\mathcal O}(\delta),
  \nonumber\\
6 \int\limits^{1-\delta}_{2\delta} dx_{1} 
\int\limits^{1-\delta}_{1-x_1+\delta} 
dx_3 \, \tilde{g}^{\rm{reg}}_7(x_1,x_3) &=& \frac{1240}{3} - 264 \zeta(2)
 + {\mathcal O}(\delta).
\label{1b.6}
\end{eqnarray}

Observing that the presence of the functions $h_6(x_1,x_3)$ and $h_7(x_1,x_3)$
does not change the singularity structure of the integrals over the variables
$x_3$, $x_1$, and $t$ in this order, the decomposition of Eq.~(\ref{1b.1})
leads to
\begin{equation}
I_i = I_i^{\rm{sing}} + I_i^{\rm{reg}},\qquad
I_i^{\rm{sing,\,reg}}=
6 \int\limits^{1-\delta}_{2\delta} dx_{3} 
\int\limits^{1-\delta}_{1-x_3+\delta}
dx_1 \, \tilde{g}^{\rm{sing,\,reg}}_i(x_1,x_3) h_i(x_1,x_3) , 
\label{1b.7}
\end{equation}
with $i=6,7$.

Our evaluation yields
\begin{eqnarray}
I_6^{\rm{sing}} &=&  9 \ln\delta + 45 
+\frac{9}{2}\zeta_2
-\frac{63}{2}\zeta_3  
+{\mathcal O}(\delta),
\label{1b.11a} \\
I_6^{\rm{reg}} &=& - \frac{422}{3} + \zeta_2 \Big( \frac{1877}{3}  
   - 1590 \logtwo - 288 \logtwo^2 \Big)
   + \frac{2719}{2}\zeta_3 - 24 \logtwo^4 + \frac{7677}{16}\zeta_4 
\nonumber \\
&&{}-576 \Li_4\left(\frac{1}{2}\right)
+\frac{35}{\sqrt{2}}G_3
+{\mathcal O}(\delta),
\label{1b.11x}\\
I_7^{\rm{sing}} &=& -9 \ln\delta - 36 - \frac{27}{2}\zeta_2
+ \frac{63}{2}\zeta_3 
+{\mathcal O}(\delta),
\label{1b.7a}\\
I_7^{\rm{reg}} &=&  297 + \zeta_2 \Big( -222  
   +486 \logtwo \Big) - \frac{567}{2}\zeta_3 
+ \frac{315}{16}\zeta_4 
+ \frac{24}{\sqrt{2}}G_3
+{\mathcal O}(\delta),
\label{1b.8}
\end{eqnarray}
where \footnote{%
The subscript of the universal contribution $G_3$ is to indicate that all its
parts carry weight three.}
\begin{eqnarray}
G_3 &=& 12 \zeta_2 \logtwo - \logtwo^3
  - 39\zeta_2\logR - 3\logtwo^2 \logR + \logR^3
  - \frac{21}{4} \zeta_3
  + 48 \Li_3\left(\frac{1}{\sqrt{2}}\right) 
\nonumber\\
&&{}+
  3 \re \left[
        \Li_3\left(\frac{1-\sqrt{2}}{2}\right)
      - \Li_3\left(\frac{1+\sqrt{2}}{2}\right) 
             \right].
\label{1b.9}
\end{eqnarray}
As can be seen from Eqs.~(\ref{1b.11a}) and (\ref{1b.7a}), $\ln\delta$
cancels in the sum $I_6+I_7$.
Here and in the following, we use the short-hand notations
\begin{equation}
  \logtwo = \ln{2}, \qquad
  \logtri = \ln{3}, \qquad
  \logR   = \ln\left(1+\sqrt{2}\right).
\label{1b.10}
\end{equation}
Furthermore,
\begin{equation}
\Si_{n,p}(x)=\frac{(-1)^{n+p-1}}{(n-1)!p!}
\int\limits_0^1\frac{dt}{t} \, \ln^p(1-tx) \, \ln^{n-1}t
\label{eq:s}
\end{equation}
is the generalized Nielsen polylogarithm, $\Li_n(x)=\Si_{n-1,1}(x)$ the
polylogarithm of order $n$, and $\zeta_n=\zeta(n)=\Li_n(1)$, with
$\zeta(x)$ being Riemann's zeta function \cite{Lewin,Devoto:1983tc}.

The result of Eq.~(\ref{1b.9}), which is the most complicated part arising from
the terms with $i=6$ and 7 in Eq.~(\ref{a.2}), assumes a rather simple form
when written as an infinite series,
\begin{equation}
\label{1b.13}
 \frac{\sqrt{2}}{3}  G_3 =
  14\zeta_3 - 24 \zeta_2 \logtwo - \frac{1}{2}
\sum\limits_{n=1}^\infty \frac{\Gamma^2(n)}{\Gamma(2n)}4^n
\left[ \psi^\prime\left(\frac{n+2}{2}\right) 
- \psi^\prime\left(\frac{n+1}{2}\right) \right],
\end{equation}
where $\psi^{(m)}(n)$ is the $(m+1)$-th logarithmic derivative of the 
$\Gamma$ function,
$\Gamma(x)=\int_0^\infty dt\,{\mathrm e}^{-t}t^{x-1}$.
We can now apply the well-known relations for $\Gamma$ and $\psi$ functions,
\begin{eqnarray}
\frac{\Gamma^2(n)}{2\Gamma(2n)} &=& 
\frac{1}{\left( 2n \atop n \right)}\, \frac{1}{n},
\label{1b.14}\\
\psi^\prime\left(\frac{n+2}{2}\right)
-\psi^\prime\left(\frac{n+1}{2}\right)
&=&(-1)^n 4 \left[ -\frac{1}{2}\zeta_2 - S_{-2}(n)\right],
\label{1b.14a}
\end{eqnarray}
where
\begin{equation}
S_{\pm m}(n) = \sum_{j=1}^n \frac{(\pm 1)^j}{j^m}
\label{1b.15}
\end{equation}
is the harmonic sum.
Using Eqs.~(\ref{1b.14}) and (\ref{1b.14a}), the constant $G_3$ is
rewritten in terms of so-called inverse central binomial sums, {\it i.e.}\
sums of the form
\begin{equation}
  \sum_{n=1}^{\infty} \frac{z^n}{\left( 2n \atop n \right)} \phi(n),
\label{ICBS}
\end{equation}
where $\phi(n)$ is some combination of harmonic sums and factors like $1/n$,
and $z$ is some number. 
Sums of such type were studied in great detail in
Refs.~\cite{Fleischer:1997bw,Fleischer:1998nb,Kalmykov:2000qe,%
Davydychev:2003mv}.

It is known that, for the series in Eq.~(\ref{ICBS}), there exists a nonlinear
transformation, 
\begin{equation}
y = \frac{\sqrt{z-4}-\sqrt{z}}{\sqrt{z-4}+\sqrt{z}},
\label{ydefinition}
\end{equation}
which leads to great simplifications in many cases.
The series in the new variable $y$ does not have a binomial coefficient and can
be summed, yielding expressions involving generalized polylogarithms
$\Si_{n,p}(y)$.

Now we can explain the appearance of the prefactor $1/\sqrt{2}$ in front of
$G_3$ in Eqs.~(\ref{1b.11x}) and (\ref{1b.8}).
Such a prefactor has not appeared in single-scale calculations so far. 
The point is that all inverse binomial series involving products of the factor 
$1/n$ and some function $f(n)$ that is a combination of the $\psi$ function and
its derivatives have the form (see, for example, Ref.~\cite{Davydychev:2003mv})
\begin{equation}
\sum\limits_{n=1}^\infty \frac{\Gamma^2(n)}{\Gamma(2n)}  z^n  f(n)
= 2 \sum\limits_{n=1}^\infty \frac{1}{\left( 2n \atop n \right)}
\, \frac{z^n}{n}  f(n) = \frac{1-y}{1+y}  F(y),
\label{1b.15a}
\end{equation}
where $F(y)$ is some combination of generalized polylogarithms  and $y$ is
defined by Eq.~(\ref{ydefinition}).
Note that Eq.~(\ref{1b.14a}) contains the binomial sum $S_{-2}(n)$, which is
related to the basic one, $S_{-2}(n-1)$, via
\begin{equation}
S_{-2}(n) = S_{-2}(n-1) + \frac{(-1)^n}{n^2}.
\label{1b.15b}
\end{equation}
Thus, the last term on the r.h.s.\ of Eq.~(\ref{1b.15b}) leads to $z=4$ in
Eq.~(\ref{1b.13}), which translates to $y=-1$ via Eq.~(\ref{ydefinition}). 
This term then cancels the term $ 14\zeta_3 - 24 \zeta_2l_2$ on the r.h.s.\ of
Eq.~(\ref{1b.13}). 
For the term $-\zeta_2/2-S_{-2}(n-1)$ on the r.h.s.\ of Eq.~(\ref{1b.14a}), we
have $z=-4$ so that the variable $y$ from Eq.~(\ref{ydefinition}) assumes the
value
\begin{equation}
r=\frac{\sqrt{2}-1}{\sqrt{2}+1}.
\label{1b.15c}
\end{equation}
This explains the appearance of the factor $1/\sqrt{2}$ in Eqs.~(\ref{1b.11x})
%,
and (\ref{1b.8}),
%, and (\ref{1b.13}), 
since $(1-y)/(1+y)=1/\sqrt{2}$.

Finally, we can rewrite Eq.~(\ref{1b.9}) as
\begin{equation}
G_3 = 21 \zeta_2 l_r  - \frac{1}{12}  l_r^3
  - 5 l_r \Li_2(r) +5 \Li_3(r) - 50 \Si_{1,2}(r)
  + 4 \Si_{1,2}(r^2)  + 34 \zeta_3,
\label{1b.16}
\end{equation}
where $l_r=\ln r$.

\section{Evaluating integrals from series}
\label{sec:some}

Let us now consider several typical integrals that arise upon the first 
integration \footnote{%
Some examples were already considered in Ref.~\cite{Kniehl:2008dt}.}.
Our first example of the remaining two-fold integrals reads
\begin{equation}
   I_{\pm} = \int\limits_0^1\frac{dt}{t}\int\limits_0^1
\frac{dx}{x} \ln[1 \mp 4t(1-t)(1-x)]\ln(1-x).
\end{equation}
%We evaluate both $I_+$ and $I_-$.
Direct integration over $t$ or $x$ would lead to rather complicated functions
in the remaining variable.
Instead, we Taylor expand the first logarithm using
$\ln(1-q)=-\sum_{n=1}^\infty q^n/n$ to obtain
\begin{equation}
I_{\pm} = - \sum\limits_{n=1}^\infty \frac{(\pm 4)^n}{n} 
         \int\limits_0^1 \frac{dt}{t} [t(1-t)]^n 
         \int\limits_0^1 \frac{dx}{x}(1-x)^n\ln(1-x).
\end{equation}
Now, the two integrals are separated and can be solved in terms of Euler's
$\Gamma$ function.
Using
\begin{equation}
   \int\limits_0^1 \frac{dx}{x}(1-x)^n\ln(1-x)=  - \psi^\prime(n+1),
\end{equation}
we finally have
\begin{equation}
  I_{\pm} = \sum\limits_{n=1}^\infty \frac{\Gamma^2(n)}{\Gamma(2n)} \, 
\frac{(\pm 4)^n}{2n} 
        \psi^\prime(n+1)
= \sum\limits_{n=1}^\infty \frac{1}{\left( 2n \atop n \right)}
\, \frac{(\pm 4)^n}{n^2} \left[ \zeta_2 - S_{2}(n)\right].
\label{Ires}
\end{equation}
Clearly, in the cases of $I_+$ and $I_-$, the argument $z$ in Eq.~(\ref{ICBS})
is equal to $4$ and $-4$, respectively.

The case of $I_+$ is simpler and leads to a smaller number of constants. 
Indeed, we can use the results of
Refs.~\cite{Kalmykov:2000qe,Davydychev:2003mv} to obtain
\begin{eqnarray}
\sum\limits_{n=1}^\infty \frac{1}{\left( 2n \atop n \right)}
\, \frac{4^n}{n^2} 
&=& 3\zeta_2,\qquad
\sum\limits_{n=1}^\infty \frac{1}{\left( 2n \atop n \right)}
\, \frac{4^n}{n^2} 
S_{2}(n-1)
= \frac{15}{4}\zeta_4,
\nonumber\\
\sum\limits_{n=1}^\infty \frac{1}{\left( 2n \atop n \right)}
\, \frac{4^n}{n^4} 
&=& 4\zeta_2l_2^2+\frac{l_2^4}{3}+8\Li_4\left(\frac{1}{2}\right)
-\frac{19}{4}\zeta_4,
\end{eqnarray}
and so on.
According to Ref.~\cite{Davydychev:2003mv}, after transformation to the 
variable $y$ of Eq.~(\ref{ydefinition}), we arrive at polylogarithms of
argument $-1$,  which are expressed in terms of alternating and
non-alternating Euler--Zagier sums, such as
$\zeta(\pm a) =\sum_{n=1}^{\infty}(\pm 1)^n/n^a$,
$\zeta(\pm a, \pm b) =\sum_{m=1}^{\infty}\sum_{n=m+1}^{\infty}
(\pm 1)^n (\pm 1)^m/(n^a m^b)$, etc.

Let us now turn to the case of $I_-$ in Eq.~(\ref{Ires}).
The argument $z=-4$ gives $y=r$
%, as may be seen from Eqs.~(\ref{ydefinition})
%and (\ref{1b.15c}) {\bf repetition from above?},
and leads to a new type of constants.
Again using formulas from Ref.~\cite{Davydychev:2003mv}, we have
\begin{eqnarray}
\sum\limits_{n=1}^\infty \frac{1}{\left( 2n \atop n \right)}
\, \frac{(- 4)^n}{n^2}
&=&  -\frac{1}{2} l_r^2, \qquad
\sum\limits_{n=1}^\infty \frac{1}{\left( 2n \atop n \right)}
\, \frac{(- 4)^n}{n^2} S_{2}(n-1)
=  \frac{1}{24} l_r^4,
\nonumber\\
\sum\limits_{n=1}^\infty \frac{1}{\left( 2n \atop n \right)}
\, \frac{(- 4)^n}{n^4}
&=&
  - \frac{2}{3}  l_r^3 - \frac{1}{8}  l_r^4
  + 4 \Si_{2,2}(r) - 4 \Li_4(r) + 4 l_2 [\Li_3(r)-\zeta_3]
\nonumber\\
&&{}+ 4 l_r [\Li_3(r)- \Si_{1,2}(r)-l_2 \Li_2(r)] 
- l_r^2 [2\Li_2(r)+l_2^2].
\end{eqnarray}
With the help of the relations listed in Appendix~\ref{sec:c}, $I_{\pm}$ can be
alternatively expressed as
\begin{eqnarray}
I_+ &=&-4\zeta_2l_2^2-\frac{l_2^4}{3}+\frac{17}{2}\zeta_4
-8\Li_4\left(\frac{1}{2}\right),
\label{5f.4}\\
I_- &=&  \zeta_4 - \frac{1}{3} l_2^4 + 2 l_2^2 \zeta_2 + 5 l_2^2 l_R^2
      -  \frac{19}{2} l_R^2 \zeta_2 - \frac{5}{3} l_R^4
\nonumber\\
&&{} - 4\logR  \re \left[
        \Li_3\left(\frac{1-\sqrt{2}}{2}\right)
      - \Li_3\left(\frac{1+\sqrt{2}}{2}\right)
             \right]
\nonumber\\
&&{}  - 4 \re \left[
        \Li_4\left(\frac{1-\sqrt{2}}{2}\right)
      + \Li_4\left(\frac{1+\sqrt{2}}{2}\right)
    \right].
\label{5f.4a}
\end{eqnarray}

It has been observed empirically that, at weight four, the terms that are not
expressed through the usual Riemann zeta function $\zeta(a)$ often come in the
combination $b_4=l_2^2(l_2^2/3 -2\zeta_2)+8\Li_4(1/2)$ introduced by
Broadhurst in Ref.~\cite{Broadhurst:1996az}.
Examples include the three-loop QCD correction to the electroweak $\rho$
parameter \cite{Avdeev:1994db}, the electron anomalous magnetic moment at
three loops \cite{Laporta:1996mq}, critical exponents in high orders of
perturbation theory \cite{Broadhurst:1996yc}, the heavy-quark contribution to
the vacuum polarization function at four loops in QCD \cite{Kniehl:2006bf},
and the matching conditions for the strong-coupling constant at five loops
\cite{Kniehl:2006bg}.
Our result for $I_+$ in Eq.~(\ref{5f.4}) exhibits a violation of this
empirical observation.
In fact, the non-zeta terms form some different combination there.

Another class of typical integrals yields sums involving $\psi$ functions of
half-integer arguments (see Appendix B), {\it e.g.}\
\begin{eqnarray}
J_{\pm}&=&\int\limits_0^1 \frac{dt}{t}\int\limits_0^1dx
     \frac{\ln[1 \mp 4t(1-t)(1-x)]\ln(1-x)}{x-2}
\nonumber\\
&=&
\sum\limits_{n=1}^\infty \frac{(\pm 4)^n}{8n}\,
\frac{\Gamma^2(n)}{\Gamma(2n)}
  \left[\psi^\prime\left(\frac{n+2}{2}\right)
  -\psi^\prime\left(\frac{n+1}{2}\right) \right]  
\nonumber\\
&=& \sum\limits_{n=1}^\infty \frac{1}{\left( 2n \atop n \right)}
\, \frac{(\mp 4)^n}{n^2} \left[ -\frac{1}{2}\zeta_2 - S_{-2}(n-1) - 
\frac{(-1)^n}{n^2}  \right].
\label{Ires1}
\end{eqnarray}
Following a similar strategy as above and using formulas from
Sec.~\ref{sec:eval}, we may express $J_{\pm}$ in terms of known irrational
constants, as
\begin{eqnarray}
J_+ &=&  -\frac{5}{2} \zeta_2 l_2^2 + \frac{17}{48} l_2^4 + 
\frac{21}{4} \zeta_4
  - 9 \zeta_2 l_2 l_R + \frac{19}{2} \zeta_2 l_R^2 
  + \frac{5}{12} l_R^4
\nonumber\\
&&{} - 9 \re \left[
        \Li_4\left(\frac{1-\sqrt{2}}{2}\right)
      + \Li_4\left(\frac{1+\sqrt{2}}{2}\right)
    \right]
+ 4 \left[
        \Li_4\left(\frac{2-\sqrt{2}}{4}\right)
      + \Li_4\left(\frac{2+\sqrt{2}}{4}\right)
    \right]
\nonumber\\
&=& -\frac{5}{2} \zeta_2 l_2^2 + \frac{17}{48} l_2^4 + 
\frac{21}{4} \zeta_4 -G_4,
\nonumber\\
J_- &=&  \frac{1}{2} \zeta_2l_2^2- \frac{49}{48} l_2^4 -\zeta_4
+6\Li_4\left(\frac{1}{2}\right) 
+  \logR^2 \left(3\logtwo^2- \frac{11}{2} \zeta_2 \right)
- \frac{7}{4} \logR^4 
\nonumber\\
&&{} +  \logR \left\{\frac{1}{3} \logtwo^3 + 5 \zeta_2\logtwo 
  + \frac{7}{4} \zeta_3
  -16 \Li_3\left(\frac{1}{\sqrt{2}}\right)
  - 5 \re \left[
        \Li_3\left(\frac{1-\sqrt{2}}{2}\right)
      - \Li_3\left(\frac{1+\sqrt{2}}{2}\right)
             \right]
\right\}
\nonumber\\
&&{}+ 5 \re \left[
        \Li_4\left(\frac{1-\sqrt{2}}{2}\right)
      + \Li_4\left(\frac{1+\sqrt{2}}{2}\right)
    \right]
- 4 \left[
        \Li_4\left(\frac{2-\sqrt{2}}{4}\right)
      + \Li_4\left(\frac{2+\sqrt{2}}{4}\right)
    \right],
\nonumber\\
\end{eqnarray}
where $G_4$, expressed with the help of the variable $r$ defined in
Eq.~(\ref{1b.15c}), is given in Eq.~(\ref{AA.5}).

These results again contain various contributions of polylogarithms with
argument $y=-1$, arising from terms of the form $(-1)^n/n^2$ on the
r.h.s.\ of Eq.~(\ref{Ires1}) for $J_+$ and terms of the form
$-\zeta_2/2 -S_2(n-1)$ on the r.h.s.\ of Eq.~(\ref{Ires1}) for $J_-$,
and with argument $y=r$, arising from the residual terms.

Unfortunately, not all integrals can be computed so straightforwardly.
In more complicated cases, the integrations are not separated after expansion
to infinite series.
We then rely on the PSLQ algorithm \cite{PSLQ}, which allows one to
reconstruct the representation of a numerical result known to very high
precision in terms of a linear combination of a set of irrational constants
with rational coefficients, if that set is known beforehand.
The experience gained with the explicit solution of the simpler integrals
helps us to exhaust the relevant sets.
In order for the PSLQ algorithm 
to work in our applications, the numerical values of the
integrals must be known up to typically 150 decimal figures.
However, for some integrals more accurate determinations are required.
The success of the application of the PSLQ algorithm also relies on the fact
that only certain combinations of polylogarithms, like $G_3$ in
Eqs.~(\ref{1b.9}) and (\ref{1b.16}), $G_4$ in Eq.~(\ref{AA.5}), and
$\tilde{G}_4$ in Eq.~(\ref{AA.4}) are incorporated as independent structures.

\section{Results}
\label{sec:res}

Finally, to get rid of complex polylogarithms, such as
$\Li_4[(1+\sqrt{2})/2]$, in the above formulas, we transform all
polylogarithms to arguments of value below unity.
To this end, we need transformation formulas through weight four.
Some of these formulas are listed in Appendix~\ref{sec:c}.
After a laborious calculation, we obtain the final result for the one-loop
correction
\begin{eqnarray}
\label{Ares}
\frac{2}{9}(\pi^2-9) A 
&=&
\frac{56}{27}
  + \frac{19}{6} \logtwo
  + \zeta_2 \left( - \frac{901}{216}
  - \frac{2701}{108} \logtwo
  + \frac{253}{24} \logtwo^2 \right)
  + \frac{11449}{432} \zeta_3
\nonumber\\
&&{}
  + \frac{59}{576} \logtwo^4
  - \frac{12983}{192} \zeta_4
  + \frac{251}{6} \Li_4\left(\frac{1}{2}\right)
  + \tilde{G}_4 +  \frac{7}{4} G_4
+ \frac{7}{6\sqrt{2}} G_3, 
\end{eqnarray}
where the constants $G_3$, $G_4$, and $\tilde{G}_4$ are specified in
Eqs.~(\ref{1b.16}), (\ref{AA.5}), and (\ref{AA.4}), respectively.
Transforming the polylogarithmic functions by means of the formulas given
in Appendix~\ref{sec:c}, we arrive at the form of Ref.~\cite{Kniehl:2008ia},
\begin{eqnarray}
\frac{2}{9}(\pi^2-9) A&=&
\frac{56}{27} 
  + \frac{19}{6} l_2
  - \frac{901}{216}\zeta_2
  - \frac{2701}{108}\zeta_2 l_2
  + \frac{11449}{432} \zeta_3
  + \frac{253}{24}\zeta_2 l_2^2
  + \frac{913}{64} \zeta_2 l_3^2
  + \frac{251}{144} l_2^4
\nonumber\\
&&{}
  + \frac{83}{256} l_3^4
  - \frac{91}{6} \zeta_3 l_2
  - \frac{11303}{192}\zeta_4
\nonumber \\
&&{} 
  - \frac{21}{4} \zeta_2 l_2 l_r
  - \frac{49}{16} \zeta_2 l_r^2
  + \frac{7}{16} l_2 l_r^3
  + \frac{35}{384} l_r^4
  - \frac{35}{8} \zeta_3 l_r
  + \frac{581}{16} \zeta_2 \Li_2\left(\frac{1}{3}\right)
\nonumber \\
&&{} 
  - \frac{21}{2} l_2 \Li_3(-r)
  - \frac{7}{2} l_r \Li_3(-r)
  + \frac{63}{4} l_2 \Li_3(r)
  + \frac{63}{8} l_r \Li_3(r)
\nonumber \\
&&{} 
  - \frac{249}{32} \Li_4\left(-\frac{1}{3}\right)
  + \frac{249}{16} \Li_4\left(\frac{1}{3}\right)
  + \frac{251}{6} \Li_4\left(\frac{1}{2}\right)
  + 7 \Li_4(-r) 
  - 7 \Si_{2,2}(-r)
\nonumber\\
&&{}
  - \frac{63}{4} \Li_4(r)
  + \frac{63}{4} \Si_{2,2}(r)
%\nonumber \\
%&&{}  
 + \frac{7}{\sqrt2}\left[ 
    \frac{7}{2} \zeta_2 l_r
  - \frac{1}{72} l_r^3
  - \frac{5}{6} l_r \Li_2(r)
\right.
\nonumber\\
&&{}  +\left. \frac{5}{6} \Li_3(r)
  - \frac{25}{3} \Si_{1,2}(r)
  + \frac{2}{3} \Si_{1,2}(r^2)
      + \frac{17}{3} \zeta_3  \right]  \,,
\label{Ares1}
\end{eqnarray}
where $r$ is given in Eq.~(\ref{1b.15c}).

%The constant $C$ in Eq.~(\ref{Gamma}) is related to $A$ through
%\cite{Kniehl:2000dh}
%\begin{equation}
%\label{Cres}
%C = - \frac{1}{3} A + \frac{229}{30} - 8 \ln{2}.
%\end{equation}
From Eqs.~(\ref{Ares1}) and (\ref{Cres}), $A$ and $C$ can be numerically
evaluated with arbitrary precision,
\begin{eqnarray}
A &=& -10.286\,614\,808\,628\,262\,240\,150\,169\,210\,991\,253\,179\,644\,%
007\,490\,228\,232\,410\dots\,,
\nonumber\\
C &=&  5.517\,027\,491\,729\,858\,271\,378\,866\,098\,665\,005\,181\,944\,%
001\,421\,860\,702\,103\,921\dots\,.\quad
\end{eqnarray}
These numbers agree with the best existing numerical evaluations
\cite{Adkins:2005eg,Kniehl:2000dh} within the quoted errors.

\section{Conclusion}
\label{sec:con}

We presented the details of our evaluation of the
${\mathcal O}(\alpha)$ and ${\mathcal O}(\alpha^3\ln\alpha)$ corrections to
the total decay width of $o$-Ps, {\it i.e.}\ of the coefficients $A$ and $C$
in Eq.~(\ref{Gamma}), respectively, which had been presented in our previous
Letter \cite{Kniehl:2008ia} in closed analytic form.
We discussed the nature and the origin of new irrational constants that appear
in the final results.
They were shown to be some particular cases of inverse central binomial sums
and corresponding generalized polylogarithms.
These constants enlarge the class of the known constants  in single-scale
problems.

The ${\mathcal O}(\alpha^2)$ correction $B$ in Eq.~(\ref{Gamma}) still remains 
analytically unknown.

\section*{Acknowledgments}

We are grateful to G.S.~Adkins for providing us with the computer code
employed for the numerical analysis in Ref.~\cite{Adkins:2005eg}.
The work of B.A.K. was supported in part by the German Federal Ministry for
Education and Research BMBF through Grant No.\ 05H09GUE.
The work of A.V.K. was supported in part by the German Research Foundation DFG
through Grant No.\ INST 152/465--1,
by the Heisenberg-Landau Program through Grant No.~5,
and by the Russian Foundation for Basic Research through Grant
No.~07--02--01046--a.
The work of O.L.V. was supported in part by the Helmholtz Association HGF
through Grant No.\ HA~101.

%%%%%%%%%%%%%%%%%%%%%%%%%%%%%%%%%%%%%%%%%%%%%%%%%%%%%%%%%%%%%%%%%%%%%%%%%%%%%%%%%%
%%%%%%%%%%%%%%%%%%%%%%%%%%%%%%%%%%%%%%%%%%%%%%%%%%%%%%%%%%%%%%%%%%%%%%%%%%%%%%%%%%
%
%           APPENDICES
%
%%%%%%%%%%%%%%%%%%%%%%%%%%%%%%%%%%%%%%%%%%%%%%%%%%%%%%%%%%%%%%%%%%%%%%%%%%%%%%%%%%
%%%%%%%%%%%%%%%%%%%%%%%%%%%%%%%%%%%%%%%%%%%%%%%%%%%%%%%%%%%%%%%%%%%%%%%%%%%%%%%%%%

\appendix

\section{Detailed results}
\label{sec:a}
\def\theequation{A\arabic{equation}}
\setcounter{equation}0

In this appendix, we present separate results for the integrals $I_i$ of
Eq.~(\ref{1b.7}) with $i=0,\ldots,7$.
Note, that not all of them a finite in the limit $\delta\to0$.
We have:
\begin{eqnarray}
I_0 &=& 204 - 142\zeta_2,
\nonumber\\ 
I_1 &=& 51 + 90 \logtwo - 228\zeta_2 + \frac{362}{3}\zeta_2\logtwo
   + \frac{1273}{12}\zeta_3,
\nonumber\\ 
I_2 &=&  - 40 - \frac{346}{5}\zeta_2 - 72\zeta_2\logtwo + 42\zeta_3
   - \frac{17}{\sqrt{2}} G_3,
\nonumber\\ 
% I_{3} + I_{5}  &=& -179 - \frac{81}{4} \pi^2 + 24 \logtwo  
% +\frac{407}{6} \pi^2 \logtwo - 305 \zeta_3
%- \frac{13163}{480}\pi^4 +46\pi^2 \logtwo^2
%+\frac{41}{4} \logtwo^4 
%\nonumber\\
%&&{}+1056 \Li_4\left(\frac{1}{2}\right) + 36 \tilde G_4 + 36 G_4,
%\nonumber\\
I_{3} &=&   144 \zeta_2\ln\delta - 59 
  + 24\logtwo 
  + \zeta_2 \left( -\frac{219}{2}+371\logtwo-294\logtwo^2 \right)
  + 52\zeta_3 - 17\logtwo^4 - \frac{17121}{16}\zeta_4
\nonumber\\
&&{}
  - 408 \Li_4\left(\frac{1}{2}\right) + 36 \tilde{G}_4,
\nonumber \\
  I_{4} &=& 
-\frac{380}{3}+ \zeta_2 \left( \frac{328}{15}
      - 252\logtwo 
      +\frac{783}{2} \logtwo^2 \right)
+ 35 \zeta_3 
+\frac{279}{16} \logtwo^4 
- \frac{1863}{4}\zeta_4
\nonumber\\
&&{}
+ 1026 \Li_4\left(\frac{1}{2}\right)
+ 27 G_4,
\nonumber\\ 
I_{5} &=&  - 144 \zeta_2\ln\delta - 120 
+ 6\zeta_2\left(-3+6\logtwo+95\logtwo^2\right)
  - 357\zeta_3 + \frac{109}{4}\logtwo^4 - 1398\zeta_4
\nonumber\\
&&{}
  + 1464 \Li_4\left(\frac{1}{2}\right) + 36 G_4,
\nonumber \\
I_{6} &=&  9 \ln\delta - \frac{287}{3} 
  + \zeta_2 \left(\frac{3781}{6} - 1590\logtwo - 288 \logtwo^2 \right)
  + 1328 \zeta_3
  - 24 \logtwo^4 
  + \frac{2559}{2}\zeta_4 
\nonumber \\
&&{}-576 \Li_4\left(\frac{1}{2}\right)
+\frac{35}{\sqrt{2}}G_3,
\nonumber\\
I_{7}  &=&  -9 \ln\delta + 261 + \zeta_2
\left(- \frac{471}{2} + 486\logtwo \right)
- 252 \zeta_3 + \frac{315}{16}\zeta_4 
+ \frac{24}{\sqrt{2}}G_3,
\label{AA.3}
%\nonumber\\
%&&{}
\end{eqnarray}
where $G_3$ is given in
Eqs.~(\ref{1b.9}) and (\ref{1b.16}), and
\begin{eqnarray}
\tilde G_4 &=&  \frac{913}{64} \zeta_2 \logtri^2
+ \frac{581}{16} \zeta_2 \Li_2\left( \frac{1}{3} \right)
  + \frac{249}{32} \left[ 2 \Li_4\left(\frac{1}{3}\right)
  -  \Li_4\left( -\frac{1}{3}\right)\right] 
+ \frac{83}{256} \logtri^4
\nonumber\\
&&{} -\frac{119}{12}  \zeta_3 \logtwo,
\label{AA.4}\\
%G_4 &=&  9\zeta_2 \logtwo \logR - \frac{19}{2} \zeta_2 \logR^2
%  - \frac{5}{12} \logR^4
%+ 9 \re \left[
%        \Li_4\left(\frac{1-\sqrt{2}}{2}\right)
%      + \Li_4\left(\frac{1+\sqrt{2}}{2}\right)
%    \right]
%\nonumber\\
%&&{}- 4 \left[
%        \Li_4\left(\frac{2-\sqrt{2}}{4}\right)
%      + \Li_4\left(\frac{2+\sqrt{2}}{4}\right)
%    \right].
%\\
G_{4} &=&  
    \frac{15}{16}\logtwo^4 
  + \frac{1}{4} \logtwo l_r^3
  + \frac{5}{96} l_r^4
  + 5 \zeta_4
  + \zeta_2 \left( - 3\logtwo l_r - \frac{7}{4} l_r^2 \right)
  + \zeta_3 \left( - 3\logtwo - \frac{5}{2} l_r \right)
\nonumber\\
&&{}
  + \left( 9\logtwo + \frac{9}{2} l_r \right) \Li_3(r) 
  + ( -6\logtwo - 2 l_r ) \Li_3(-r) 
  - 9 \left[ \Li_4(r) - {\rm S}_{2,2}(r) \right] 
\nonumber\\
&&{}
  + 4 \left[ \Li_4(-r) - {\rm S}_{2,2}(-r) \right]. 
\label{AA.5}
\end{eqnarray}

  From Eqs.~(\ref{AA.3}) it is clear that $\ln\delta$ cancels
in the sum $\sum_{j=0}^7 I_j$.

\boldmath
\section{Expansions of $\Gamma$ and $\psi$ functions about half-integer 
   arguments}
\label{sec:b}
\unboldmath 
\def\theequation{B\arabic{equation}}
\setcounter{equation}0

In this appendix, we present some useful relations between derivatives of the
$\psi$ function with half-integer arguments and the $\psi$ and $\beta$
functions with integer arguments, and consider the expansion of the $\Gamma$
function in the vicinity of half-integer arguments.

Starting from the well-known relations between the $\psi$ and $\beta$
functions,
\begin{eqnarray}
\psi(2z) &=& \frac{1}{2} \left[\psi\left(z+\frac{1}{2}\right)+\psi(z)\right]
+ \logtwo,
\nonumber \\
\beta(2z) &=& 
\frac{1}{2} \left[\psi\left(z+\frac{1}{2}\right)-\psi(z) \right],
\label{B.1} 
\end{eqnarray}
and differentiating them $m$ ($m>0$) times, we have  
\begin{eqnarray}
2^{m+1} \psi^{(m)}(2z) &=& 
\psi^{(m)}\left(z+\frac{1}{2}\right)+\psi^{(m)}(z),
\nonumber \\
2^{m+1} \beta^{(m)}(2z) &=& 
\psi^{(m)}\left(z+\frac{1}{2}\right)-\psi^{(m)}(z),
\label{B.2} 
\end{eqnarray}
where $\psi^{(m)}(z)$ denotes the $m$-th derivative of $\psi(z)$ etc.
We can combine Eqs.~(\ref{B.1}) and (\ref{B.2}) as
\begin{eqnarray}
\psi^{(m)}(z) &=&
2^m \left[ \psi^{(m)}(2z) - \beta^{(m)}(2z) \right] -\delta_{0m} \logtwo, 
\nonumber \\
\psi^{(m)}\left(z+\frac{1}{2}\right) &=&
2^m \left[ \psi^{(m)}(2z) + \beta^{(m)}(2z) \right]  -\delta_{0m} \logtwo,
\label{B.3} 
\end{eqnarray}
where $\delta_{mn}$ is the Kronecker symbol.

Using the series representations of the $\psi$ and $\beta$ functions
\cite{GraRy},
\begin{eqnarray}
\psi(z) &=& \psi(1) + (z-1) \sum_{k=0}^{\infty} \frac{1}{(k+1)(k+z)},
\nonumber\\
\beta(z) &=& \sum_{k=0}^{\infty} \frac{(-1)^k}{k+z},
\label{B.3b}
\end{eqnarray}
we obtain the following relations: 
\begin{eqnarray}
\psi(n+1) &=&  \psi(1) + \Si_1(n), 
\nonumber\\ 
\psi^{(m)}(n+1) &=& (-1)^m m! [
\Si_{m+1}(n) - \zeta_{m+1} ],
\nonumber \\
\beta(n+1) &=& (-1)^n [l_2+ \Si_{-1}(n)],
\nonumber\\
 \beta^{(m)}(n+1) &=& 
(-1)^{m+n} m! [
  \Si_{-(m+1)}(n) - \Si_{-(m+1)}(\infty) ],
\label{B.4} 
\end{eqnarray}
where $\Si_m(n)$ is defined in Eq.~(\ref{1b.15}).

Thus, Eqs.~(\ref{B.3}) and (\ref{B.4}) lead to the following 
results for the ``sums'' $S_m$ with half-integer arguments \footnote{%
We may use Eq.~(\ref{B.4}) as definition of the ``sums'' $S_m$  with
half-integer arguments.}:
\begin{eqnarray}
S_1\left(\frac{n}{2}\right) &=& S_1(n)+ (-1)^n S_{-1}(n) 
-[1-(-1)^n] l_2, \nonumber \\ 
S_m\left(\frac{n}{2}\right) &=& 2^{m-1} [S_m(n)+ (-1)^n S_{-m}(n)]  
+[1-(-1)^n](1-2^{m-1}) \zeta_m \qquad (m\geq 2).\
\label{B.7}
\end{eqnarray}
These equations are useful for expansions of the $\Gamma$ function in the
vicinities of half-integer arguments.
Indeed, using a well-known formula for the expansions of the $\Gamma$ function
about integer values, which was used, {\it e.g.}, in
Ref.~\cite{Kazakov:1987jk},
\begin{equation}
\frac{\Gamma(n+1+\delta)}{n!\Gamma(1+\delta)} = 
\exp \left[- \sum_{k=1}^{\infty}
\frac{1}{k} S_k(n) (-\delta)^k \right],
 \label{B.11}
\end{equation}
where $\gamma_E$ is Euler's constant, we find the corresponding 
expansions about half-integer values to be
\begin{eqnarray}
 \frac{\Gamma(n/2+1+\delta)}{\Gamma(n/2+1)\Gamma(1+\delta)} &=& 
\exp \left[-\sum_{k=1}^{\infty}
\frac{1}{k} S_k\left(\frac{n}{2}\right) (-\delta)^k \right],
\label{B.12}
\end{eqnarray}
where $S_m(n/2)$ is given by Eq.~(\ref{B.7}).
Such expansions are useful in many applications, including those in
Ref.~\cite{Kalmykov:2008ge} and references cited therein.

\section{Transformations of polylogarithms of weight four}
\label{sec:c}
\def\theequation{C\arabic{equation}}
\setcounter{equation}0

In this appendix, we present relations between the generalized polylogarithms
${\Si}_{a,b}$ of weight four ($a+b=4$) with different arguments.
Transformations at lower weights can be found in the literature
\cite{Devoto:1983tc}.
Although the derivation of these formulas is straightforward, we present them
here for the convenience of interested readers.
At weight four, there are three independent Nielsen polylogarithms, which we
choose to be $\Li_4$, $\Si_{1,3}$, and $\Si_{2,2}$.

{\bf 1.} Relations for the functions with arguments $1-y$ and $y$:
\begin{eqnarray}
\Li_{4}(1-y) &=& \zeta_4 - \Si_{1,3}(y) + \ln(1-y) [
\zeta_3 - \Si_{1,2}(y) ] + \frac{1}{2}  \ln^2(1-y) [
\zeta_2 - \Li_{2}(y)] 
\nonumber \\
&&{} - \frac{1}{6}  \ln^3(1-y)  \ln y, 
\nonumber\\ 
\Si_{2,2}(1-y) &=& \frac{1}{4}  \zeta_4 - \Si_{2,2}(y) + \ln y  
\Si_{1,2}(y)  + \ln(1-y) [
\zeta_3 - \Li_{3}(y) +  \ln y  \Li_{2}(y) ] 
\nonumber \\ 
&&{} 
+ \frac{1}{4}  \ln^2(1-y)  \ln^2 y, 
\nonumber \\ 
\Si_{1,3}(1-y) &=& \zeta_4 - \Li_{4}(y) + \ln y  
\Li_{3}(y)  - \frac{1}{2}  \ln^2 y  \Li_{2}(y) 
- \frac{1}{6} \ln^3 y \ln(1-y) . 
\label{app1}
\end{eqnarray}

{\bf 2.} Relations for the functions with arguments $-1/y$ and $-y$:
\begin{eqnarray}
\Li_{4}\left(-\frac{1}{y}\right) &=&
 - \Li_{4}(-y) -   \frac{7}{4}  \zeta_4 - \frac{1}{2}  
\zeta_2  \ln^2 y  - \frac{1}{24}  
\ln^4 y , 
\nonumber \\
\Si_{2,2}\left(-\frac{1}{y}\right) &=&
 \Si_{2,2}(-y) -2 \Li_{4}(-y) 
-  \frac{7}{4}  \zeta_4
-  \ln y  [
\zeta_3 - \Li_{3}(-y) ] 
+ \frac{1}{24}   \ln^4 y ,
\nonumber \\
\Si_{1,3}\left(-\frac{1}{y}\right) &=& 
- \Si_{1,3}(-y) +  \Si_{2,2}(-y) - \Li_{4}(-y)
- \zeta_4
-  \ln y   [ \Si_{1,2}(-y)
 - \Li_{3}(-y) ] 
\nonumber \\
&&{} -  \frac{1}{2} \ln^2 y   \Li_{2}(-y)
- \frac{1}{24}   \ln^4 y . 
\label{app2} 
\end{eqnarray}

{\bf 3.} Relations for the functions with arguments $(y-1)/y$ and $y$:
\begin{eqnarray}
\Li_{4}\left(\frac{y-1}{y}\right) &=& \Li_{4}(y) + \Si_{1,3}(y) - 
\Si_{2,2}(y) -
\frac{7}{4}  \zeta_4 + \ln(1-y) [
\Si_{1,2}(y) - \Li_{3}(y) ] 
\nonumber \\
&&{} + \frac{1}{2}  \ln^2(1-y)  \Li_{2}(y)
-   \frac{1}{2} \zeta_2  \ln^2\frac{1-y}{y}  
+   \frac{1}{24}  \ln^4(1-y) 
\nonumber \\
&&{} - \frac{1}{24}  \ln^4\frac{1-y}{y} , 
\nonumber \\ 
\Si_{2,2}\left(\frac{y-1}{y}\right) &=& 2\Li_{4}(y) - \Si_{2,2}(y) -
\frac{7}{4}  \zeta_4 
+ \ln y  \Si_{1,2}(y)  + \ln\frac{1-y}{y} 
\zeta_3 - \ln[(1-y)y] \Li_{3}(y)
\nonumber \\
&&{} +  \ln y\ln(1-y)    \Li_{2}(y)
+ \frac{1}{4} \ln^2 y \ln^2(1-y)   - \frac{1}{6} \ln^3 y \ln(1-y)   
\nonumber \\
&&{} 
+   \frac{1}{24}  \ln^4 y , 
\nonumber \\ 
\Si_{1,3}\left(\frac{y-1}{y}\right) &=& \Li_{4}(y) - \zeta_4 -  
\ln y  \Li_{3}(y)  + \frac{1}{2}  \ln^2 y  \Li_{2}(y) 
+ \frac{1}{6} \ln^3 y \ln(1-y)   
\nonumber \\
&&{} -\frac{1}{24} \ln^4 y . 
\label{app3}
\end{eqnarray}

{\bf 4.} Relations for the functions with arguments $y/(y-1)$ and $y$:
\begin{eqnarray}
\Li_{4}\left(\frac{y}{y-1}\right) &=&  \Si_{2,2}(y) -\Li_{4}(y) - \Si_{1,3}(y)
+ \ln(1-y) [\Li_{3}(y) -\Si_{1,2}(y) ] 
\nonumber \\
&&{}  - \frac{1}{2}  \ln^2(1-y)  \Li_{2}(y)
 - \frac{1}{24}    \ln^4(1-y) , 
\nonumber \\ 
\Si_{2,2}\left(\frac{y}{y-1}\right) &=& \Si_{2,2}(-y) - 2 \Si_{1,3}(y)
- \ln (1-y) \Si_{1,2}(y)  + \frac{1}{24}    \ln^4(1-y) ,
\nonumber \\ 
\Si_{1,3}\left(\frac{y}{y-1}\right) &=& - \Si_{1,3}(y) - 
\frac{1}{24}    \ln^4(1-y) .
\label{app4}
\end{eqnarray}

{\bf 5.} Relations for the functions with arguments $1/(1+y)$ and $-y$:
\begin{eqnarray}
\Li_{4}\left(\frac{1}{1+y}\right) &=& \Si_{1,3}(-y) + \zeta_4 
+ \ln(1+y)  [\Si_{1,2}(-y) -  \zeta_3 ] 
 + \frac{1}{2}  \ln^2(1+y) [\Li_{2}(-y)+  \zeta_2 ] 
\nonumber \\
&&{}  
 +  \frac{1}{6}  \ln^3(1+y)  \ln y 
-   \frac{1}{24}  \ln^4(1+y) , 
\nonumber \\ 
\Si_{2,2}\left(\frac{1}{1+y}\right) &=& 2 \Si_{1,3}(-y) - \Si_{2,2}(-y) +
\frac{1}{4}  \zeta_4 + 
+ \ln [y(1+y)]  \Si_{1,2}(-y)  
\nonumber \\
&&{} - \ln(1+y) [\Li_{3}(-y) + \zeta_3 ] +  \ln(1+y)  \ln y  \Li_{2}(-y)
+ \frac{1}{4}  \ln^2(1+y)  \ln^2 y 
\nonumber \\
&&{} - \frac{1}{6}  \ln^3(1+y)  \ln y  +   \frac{1}{24}  \ln^4 (1+y) , 
\nonumber \\ 
\Si_{1,3}\left(\frac{1}{1+y}\right) &=&   \Si_{1,3}(-y) - 
\Si_{2,2}(-y) + \Li_{4}(-y) + \zeta_4 +  
\ln y [\Si_{1,2}(-y) -
\Li_{3}(-y)  ] 
\nonumber \\
&&{} + \frac{1}{2}  \ln^2 y  \Li_{2}(-y) 
+\frac{1}{24}  \ln^4 y
-\frac{1}{24}  \ln^4\frac{1+y}{y} . 
\label{app5}
\end{eqnarray}
Equations~(\ref{app1}) and (\ref{app2}) were directly obtained from
Ref.~\cite{Devoto:1983tc}, where they are presented for the generalized
polylogarithms $\Si_{a,b}$ with arbitrary values of $a$ and $b$,
but in some complicated form less convenient for applications.
Equations~(\ref{app3})--(\ref{app5}) were found by iterated application 
of Eqs.~(\ref{app1}) and (\ref{app2}) and equations from
Ref.~\cite{Devoto:1983tc}.
They are simple and useful for applications together with equations for
$\Si_{a,b}$ from Ref.~\cite{Devoto:1983tc}, with the constraints $a+b=2$ or
$a+b=3$.

%%%%%%%%%%%%%%%%%%%%%%%%%%%%%%%%%%%%%%%%%%%%%%%%%%%%%%%%%%%%%%%%%

%%%%%%%%%%%%%%%%%%%%%%%%%%%%%%%%%%%%%%%%%%%%%%%%%%%%%%%%%%%%%%%%%%%%%%%%%%%%%%%%%%%
%%%%%%%%%%%%%  L I T E R A T U R E
%%%%%%%%%%%%%%%%%%%%%%%%%%%%%%%%%%%%%%%%%%%%%%%%%%%%%%%%%%%%%%%%%%%%%%%%%%%%%%%%%%%


\begin{thebibliography}{99}

\bibitem{Deutsch:1951zza}
  M.~Deutsch,
  %``Evidence For The Formation Of Positronium In Gases,''
  Phys.\ Rev.\  {\bf 82}, 455 (1951).
  %%CITATION = PHRVA,82,455;%%

\bibitem{Caswell:1985ui}
  W.~E.~Caswell and G.~P.~Lepage,
  %``Effective Lagrangians For Bound State Problems In QED, QCD, And Other Field
  %Theories,''
  Phys.\ Lett.\  B {\bf 167}, 437 (1986).

\bibitem{Deutsch:1951zz}
  M.~Deutsch,
  %``Three-Quantum Decay of Positronium,''
  Phys.\ Rev.\  {\bf 83}, 866 (1951).
  %%CITATION = PHRVA,83,866;%%

\bibitem{Ore:1949te}
  A.~Ore and J.~L.~Powell,
  %``Three Photon Annihilation Of An Electron - Positron Pair,''
  Phys.\ Rev.\  {\bf 75}, 1696 (1949).

\bibitem{BH}
R.~H.~Beers and V.~W.~Hughes,
Bull.\ Am.\ Phys.\ Soc.\ {\bf 13}, 633 (1968).

\bibitem{Caswell:1976nx}
  W.~E.~Caswell, G.~P.~Lepage, and J.~R.~Sapirstein,
  %``Order (Alpha) Corrections To The Decay Rate Of Orthopositronium,''
  Phys.\ Rev.\ Lett.\ {\bf 38}, 488 (1977).

\bibitem{Harris:1957zz}
  I.~Harris and L.~M.~Brown,
  %``Radiative Corrections to Pair Annihilation,''
  Phys.\ Rev.\  {\bf 105}, 1656 (1957).
  %%CITATION = PHRVA,105,1656;%%

\bibitem{Westbrook:1987zz}
  C.~I.~Westbrook, D.~W.~Gidley, R.~S.~Conti, and A.~Rich,
  %``New Precision Measurement Of The Orthopositronium Decay Rate: A Discrepancy
  %With Theory,''
  Phys.\ Rev.\ Lett.\  {\bf 58}, 1328 (1987);
  %%CITATION = PRLTA,58,1328;%%
{\bf 58}, 2153(E) (1987);
%\bibitem{Westbrook:1989zz}
%  C.~I.~Westbrook, D.~W.~Gidley, R.~S.~Conti and A.~Rich,
  %``Precision measurement of the orthopositronium vacuum decay rate using the
  %gas technique,''
  Phys.\ Rev.\  A {\bf 40}, 5489 (1989).
  %%CITATION = PHRVA,A40,5489;%%

\bibitem{Vallery:2003iz}
  R.~S.~Vallery, P.~W.~Zitzewitz, and D.~W.~Gidley,
  %``Resolution Of The Orthopositronium Lifetime Puzzle,''
  Phys.\ Rev.\ Lett.\ {\bf 90}, 203402 (2003).

\bibitem{Jinnouchi:2003hr}
  O.~Jinnouchi, S.~Asai, and T.~Kobayashi,
  %``Precision measurement of orthopositronium decay rate using SiO-2  powder,''
  Phys.\ Lett.\  B {\bf 572}, 117 (2003)
  [arXiv:hep-ex/0308030].
  %%CITATION = PHLTA,B572,117;%%

\bibitem{Caswell:1978vz}
  W.~E.~Caswell and G.~P.~Lepage,
  %``O (Alpha**2 Log (Alpha**-1)) Corrections In Positronium: Hyperfine
  %Splitting And Decay Rate,''
  Phys.\ Rev.\  A {\bf 20}, 36 (1979).

\bibitem{Khriplovich:1990eh}
  I.~B.~Khriplovich and A.~S.~Yelkhovsky,
  %``On the radiative corrections alpha**2 in alpha to the positronium decay
  %rate,''
  Phys.\ Lett.\  B {\bf 246}, 520 (1990).

\bibitem{Kar}
S. G. Karshenboim,
Sov.\ Phys.\ JETP {\bf76}, 541 (1993)
[Zh.\ Eksp.\ Teor.\ Fiz.\ {\bf103}, 1105 (1993)].

\bibitem{Stroscio:1974zz}
  M.~A.~Stroscio and J.~M.~Holt,
  %``Radiative corrections to the decay rate of orthopositronium,''
  Phys.\ Rev.\  A {\bf 10}, 749 (1974);
  %%CITATION = PHRVA,A10,749;%%
%\bibitem{Stroscio:1975fa}
  M.~A.~Stroscio,
  %``Positronium: Review Of The Theory,''
  Phys.\ Rept.\  {\bf 22}, 215 (1975);
%\bibitem{Adkins:1982zn}
  G.~S.~Adkins,
  %``Radiative Corrections To Positronium Decay,''
  Ann.\ Phys.\ (N.Y.) {\bf 146}, 78 (1983);
%\bibitem{Adkins:1992zza}
  G.~S.~Adkins, A.~A.~Salahuddin, and K.~E.~Schalm,
  %``Order-alpha corrections to the decay rate of orthopositronium in the
  %Fried-Yennie gauge,''
  Phys.\ Rev.\  A {\bf 45}, 7774 (1992);
  %%CITATION = PHRVA,A45,7774;%%
%\bibitem{Adkins:2005eg}
  G.~S.~Adkins,
  %``Analytic evaluation of the amplitudes for orthopositronium decay to  three
  %photons to one-loop order,''
  Phys.\ Rev.\ Lett.\  {\bf 76}, 4903 (1996)
  [arXiv:hep-ph/0506213].
  %%CITATION = PRLTA,76,4903;%%

\bibitem{Adkins:2000fg}
  G.~S.~Adkins, R.~N.~Fell, and J.~R.~Sapirstein,
  %``Order alpha**2 corrections to the decay rate of orthopositronium,''
  Phys.\ Rev.\ Lett.\ {\bf 84}, 5086 (2000)
  [arXiv:hep-ph/0003028];
%\bibitem{Adkins:2001zz}
%  G.~S.~Adkins, R.~N.~Fell and J.~Sapirstein,
  %``Light-by-light scattering contributions to positronium decay rates,''
  Phys.\ Rev.\  A {\bf 63}, 032511 (2001).
  %%CITATION = PHRVA,A63,032511;%%

\bibitem{Adkins:2005eg}
  G.~S.~Adkins,
  %``Analytic evaluation of the amplitudes for orthopositronium decay to  three
  %photons to one-loop order,''
  Phys.\ Rev.\ A {\bf 72}, 032501 (2005).

\bibitem{Kniehl:2000dh}
  B.~A.~Kniehl and A.~A.~Penin,
  %``Order alpha**3 ln(1/alpha) corrections to positronium decays,''
  Phys.\ Rev.\ Lett.\ {\bf 85}, 1210 (2000);
  {\bf 85}, 3065(E) (2000)
  [arXiv:hep-ph/0004267];
%\bibitem{Hill:2000qi}
  R.~J.~Hill and G.~P.~Lepage,
  %``O(alpha**2 Gamma alpha**3 gamma) binding effects in orthopositronium
  %decay,''
  Phys.\ Rev.\  D {\bf 62}, 111301(R) (2000)
  [arXiv:hep-ph/0003277];
%\bibitem{Melnikov:2000fi}
  K.~Melnikov and A.~Yelkhovsky,
  %``O(alpha**3 ln(alpha)) corrections to positronium decay rates,''
%  Phys.\ Rev.\  D 
{\it ibid.}\ {\bf 62}, 116003 (2000)
  [arXiv:hep-ph/0008099].

\bibitem{AFS}
  G.~S.~Adkins, R.~N.~Fell, and J.~R.~Sapirstein,
  Ann.\ Phys.\ (N.Y.) {\bf 295}, 136 (2002);
%\bibitem{Sillou:2004js}
  D.~Sillou,
  %``Status of orthopositronium decay rate measurements,''
  Int.\ J.\ Mod.\ Phys.\  A {\bf 19}, 3919 (2004).
  %%CITATION = IMPAE,A19,3919;%%

\bibitem{Rubbia:2004ix}
  A.~Rubbia,
  %``Positronium as a probe for new physics beyond the standard model,''
  Int.\ J.\ Mod.\ Phys.\  A {\bf 19}, 3961 (2004)
  [arXiv:hep-ph/0402151];
%\bibitem{Gninenko:2006sz}
  S.~N.~Gninenko, N.~V.~Krasnikov, V.~A.~Matveev, and A.~Rubbia,
  %``Some Aspects Of Positronium Physics,''
  Phys.\ Part.\ Nucl.\  {\bf 37}, 321 (2006);
  %%CITATION = PPNUE,37,321;%%
%\bibitem{Gninenko:2008yq}
  S.~N.~Gninenko, N.~V.~Krasnikov, and V.~A.~Matveev,
  %``Invisible $Z\prime$ as a probe of extra dimensions at the CERN LHC,''
  Phys.\ Rev.\  D {\bf 78}, 097701 (2008)
  [arXiv:0811.0974 [hep-ph]].
  %%CITATION = PHRVA,D78,097701;%%

\bibitem{Akama:1982jy}
  K.~Akama,
  %``An early proposal of 'brane world',''
  Lect.\ Notes Phys.\  {\bf 176}, 267 (1983)
  [arXiv:hep-th/0001113];
  %%CITATION = LNPHA,176,267;%%
%\bibitem{Rubakov:1983bb}
  V.~A.~Rubakov and M.~E.~Shaposhnikov,
  %``Do We Live Inside A Domain Wall?,''
  Phys.\ Lett.\  B {\bf 125}, 136 (1983);
  %%CITATION = PHLTA,B125,136;%%
%\bibitem{Visser:1985qm}
  M.~Visser,
  %``An Exotic Class Of Kaluza-Klein Models,''
%  Phys.\ Lett.\  B 
{\it ibid.}\ {\bf 159}, 22 (1985)
  [arXiv:hep-th/9910093];
  %%CITATION = PHLTA,B159,22;%%
%\bibitem{Antoniadis:1990ew}
  I.~Antoniadis,
  %``A Possible new dimension at a few TeV,''
%  Phys.\ Lett.\  B 
{\it ibid.}\ {\bf 246}, 377 (1990);
  %%CITATION = PHLTA,B246,377;%%
%\bibitem{Krasnikov:1991dt}
  N.~V.~Krasnikov,
  %``Ultraviolet fixed point behavior of the five-dimensional Yang-Mills theory,
  %the gauge hierarchy problem and a possible new dimension at the TeV scale,''
%  Phys.\ Lett.\  B 
{\it ibid.}\ {\bf 273}, 246 (1991).
  %%CITATION = PHLTA,B273,246;%%

\bibitem{ArkaniHamed:1998rs}
  N.~Arkani-Hamed, S.~Dimopoulos, and G.~R.~Dvali,
  %``The hierarchy problem and new dimensions at a millimeter,''
  Phys.\ Lett.\  B {\bf 429}, 263 (1998)
  [arXiv:hep-ph/9803315];
  %%CITATION = PHLTA,B429,263;%%
%\bibitem{Antoniadis:1998ig}
  I.~Antoniadis, N.~Arkani-Hamed, S.~Dimopoulos, and G.~R.~Dvali,
  %``New dimensions at a millimeter to a Fermi and superstrings at a TeV,''
%  Phys.\ Lett.\  B 
{\it ibid.}\ {\bf 436}, 257 (1998)
  [arXiv:hep-ph/9804398];
  %%CITATION = PHLTA,B436,257;%%
%\bibitem{Randall:1999ee}
  L.~Randall and R.~Sundrum,
  %``A large mass hierarchy from a small extra dimension,''
  Phys.\ Rev.\ Lett.\  {\bf 83}, 3370 (1999)
  [arXiv:hep-ph/9905221];
  %%CITATION = PRLTA,83,3370;%%
%\bibitem{Randall:1999vf}
%  L.~Randall and R.~Sundrum,
  %``An alternative to compactification,''
%  Phys.\ Rev.\ Lett.\  
{\it ibid.}\ {\bf 83}, 4690 (1999)
  [arXiv:hep-th/9906064].
  %%CITATION = PRLTA,83,4690;%%

\bibitem{Rubakov:2001kp}
  V.~A.~Rubakov,
  %``Large and infinite extra dimensions: An introduction,''
  Phys.\ Usp.\  {\bf 44}, 871 (2001)
  [Usp.\ Fiz.\ Nauk {\bf 171}, 913 (2001)]
  [arXiv:hep-ph/0104152].
  %%CITATION = UFNAA,171,913;%%

\bibitem{Dubovsky:2000am}
  S.~L.~Dubovsky, V.~A.~Rubakov, and P.~G.~Tinyakov,
  %``Brane world: Disappearing massive matter,''
  Phys.\ Rev.\  D {\bf 62}, 105011 (2000)
  [arXiv:hep-th/0006046].
  %%CITATION = PHRVA,D62,105011;%%

\bibitem{Oda:2000zc}
  I.~Oda,
  %``Localization of matters on a string-like defect,''
  Phys.\ Lett.\  B {\bf 496}, 113 (2000)
  [arXiv:hep-th/0006203].
  %%CITATION = PHLTA,B496,113;%%

\bibitem{Dubovsky:2000av}
  S.~L.~Dubovsky, V.~A.~Rubakov, and P.~G.~Tinyakov,
  %``Is the electric charge conserved in brane world?,''
  JHEP {\bf 0008}, 041 (2000)
  [arXiv:hep-ph/0007179];
  %%CITATION = JHEPA,0008,041;%%
%\bibitem{Dubovsky:2002xv}
  S.~L.~Dubovsky and V.~A.~Rubakov,
  %``On electric charge non-conservation in brane world,''
in {\it Proceedings of the XXXVIIth Rencontres de Moriond: 2002 Electroweak
Interactions and Unified Theories}, edited by J. Tr\^an Thanh V\^an
(Th{\^e} Gi\'oi Publishers, Vietnam, 2003), p.~367
  [arXiv:hep-th/0204205].
  %%CITATION = HEP-TH/0204205;%%

\bibitem{Gninenko:2003nx}
  S.~N.~Gninenko, N.~V.~Krasnikov, and A.~Rubbia,
  %``Extra dimensions and invisible decay of orthopositronium,''
  Phys.\ Rev.\  D {\bf 67}, 075012 (2003)
  [arXiv:hep-ph/0302205].
  %%CITATION = PHRVA,D67,075012;%%

\bibitem{Badertscher:2006fm}
  A.~Badertscher, P.~Crivelli, W.~Fetscher, U.~Gendotti, S.~Gninenko,
 V.~Postoev, A.~Rubbia, V.~Samoylenko, and D.~Sillou, 
  %``An Improved Limit on Invisible Decays of Positronium,''
  Phys.\ Rev.\  D {\bf 75}, 032004 (2007)
  [arXiv:hep-ex/0609059].
  %%CITATION = PHRVA,D75,032004;%%

\bibitem{Stroscio:1982wj}
  M.~A.~Stroscio,
  %``Exact First Order Electron Selfenergy Contribution To The Decay Rate Of
  %Orthopositronium,''
  Phys.\ Rev.\ Lett.\  {\bf 48}, 571 (1982);
%\bibitem{Adkins:1982hv}
  G.~S.~Adkins,
  %``Analytic Evaluation Of An O (Alpha) Vertex Correction To The Decay Rate Of
  %Orthopositronium,''
  Phys.\ Rev.\  A {\bf 27}, 530 (1983);
  %%CITATION = PHRVA,A27,530;%%

\bibitem{Kniehl:2008ia}
  B.~A.~Kniehl, A.~V.~Kotikov, and O.~L.~Veretin,
  %``Orthopositronium lifetime: analytic results in O(alpha) and O(alpha^3
  %ln(alpha)),''
  Phys.\ Rev.\ Lett.\  {\bf 101}, 193401 (2008)
  [arXiv:0806.4927 [hep-ph]].
  %%CITATION = PRLTA,101,193401;%%

\bibitem{Lewin}
L. Lewin,
{\it Polylogarithms and Associated Functions}
(Elsevier, New York, 1981).

\bibitem{Kniehl:2005yc}
  B.~A.~Kniehl and A.~V.~Kotikov,
  %``Calculating four-loop tadpoles with one non-zero mass,''
  Phys.\ Lett.\  B {\bf 638}, 531 (2006)
  [arXiv:hep-ph/0508238].

\bibitem{Devoto:1983tc}
  A.~Devoto and D.~W.~Duke,
  %``Table Of Integrals And Formulae For Feynman Diagram Calculations,''
  Riv.\ Nuovo Cim.\  {\bf 7N6}, 1 (1984).

\bibitem{Fleischer:1997bw}
  J.~Fleischer, A.~V.~Kotikov, and O.~L.~Veretin,
  %``The differential equation method: Calculation of vertex-type diagrams  with
  %one non-zero mass,''
  Phys.\ Lett.\  B {\bf 417}, 163 (1998)
  [arXiv:hep-ph/9707492];
%\bibitem{Fleischer:1999hp}
  J.~Fleischer, M.~Yu.~Kalmykov, and A.~V.~Kotikov,
  %``Two-loop self-energy master integrals on shell,''
%  Phys.\ Lett.\  B 
{\it ibid.}\ {\bf 462}, 169 (1999)
  [arXiv:hep-ph/9905249];
  %%CITATION = PHLTA,B462,169;%%
%\bibitem{Kotikov:2007vr}
  A.~Kotikov, J.~H.~K\"uhn, and O.~Veretin,
  %``Two-loop formfactors in theories with mass gap and Z-boson production,''
  Nucl.\ Phys.\ {\bf B788}, 47 (2008)
  [arXiv:hep-ph/0703013].

\bibitem{Fleischer:1998nb}
  J.~Fleischer, A.~V.~Kotikov, and O.~L.~Veretin,
  %``Analytic two-loop results for selfenergy- and vertex-type diagrams  with
  %one non-zero mass,''
  Nucl.\ Phys.\ {\bf B547}, 343 (1999)
  [arXiv:hep-ph/9808242].

\bibitem{Kalmykov:2000qe}
  M.~Yu.~Kalmykov and O.~Veretin,
  %``Single-scale diagrams and multiple binomial sums,''
  Phys.\ Lett.\  B {\bf 483}, 315 (2000)
  [arXiv:hep-th/0004010].

\bibitem{Davydychev:2003mv}
  A.~I.~Davydychev and M.~Yu.~Kalmykov,
  %``Massive Feynman diagrams and inverse binomial sums,''
  Nucl.\ Phys.\ {\bf B699}, 3 (2004)
  [arXiv:hep-th/0303162].
  %%CITATION = NUPHA,B699,3;%%

\bibitem{Kniehl:2008dt}
  B.~A.~Kniehl, A.~V.~Kotikov and, O.~L.~Veretin,
  %``Irrational constants in positronium decays,''
Nucl.\ Phys.\ B (Proc.\ Suppl.) {\bf 183}, 14 (2008)
  [arXiv:0811.0306 [hep-ph]].
  %%CITATION = ARXIV:0811.0306;%%

\bibitem{Broadhurst:1996az}
  D.~J.~Broadhurst,
  %``On the enumeration of irreducible k-fold Euler sums and their roles in knot
  %theory and field theory,''
  arXiv:hep-th/9604128;
  %%CITATION = HEP-TH/9604128;%%
%\bibitem{Broadhurst:1998rz}
%  D.~J.~Broadhurst,
  %``Massive 3-loop Feynman diagrams reducible to SC* primitives of  algebras of
  %the sixth root of unity,''
  Eur.\ Phys.\ J.\  C {\bf 8}, 311 (1999)
  [arXiv:hep-th/9803091].
  %%CITATION = EPHJA,C8,311;%%

\bibitem{Avdeev:1994db}
  L.~Avdeev, J.~Fleischer, S.~Mikhailov, and O.~Tarasov,
  %``0 (alpha alpha-s**2) correction to the electroweak rho parameter,''
  Phys.\ Lett.\  B {\bf 336}, 560 (1994)
  [arXiv:hep-ph/9406363];
{\bf 349}, 597(E) (1995);
  %%CITATION = PHLTA,B336,560;%%
%\bibitem{Fleischer:1994dc}
  J.~Fleischer and O.~V.~Tarasov,
  %``Application of conformal mapping and Pade approximants (omega P's) to  the
  %calculation of various two-loop Feynman diagrams,''
  Nucl.\ Phys.\ B (Proc.\ Suppl.) {\bf 37}, 115 (1994)
  [arXiv:hep-ph/9407235];
  %%CITATION = NUPHZ,37B,115;%%
%\bibitem{Chetyrkin:1995ix}
  K.~G.~Chetyrkin, J.~H.~K\"uhn, and M.~Steinhauser,
  %``Corrections of order O (G(F) M(t)**2 alpha-s**2) to the rho parameter,''
  Phys.\ Lett.\  B {\bf 351}, 331 (1995)
  [arXiv:hep-ph/9502291].
  %%CITATION = PHLTA,B351,331;%%

\bibitem{Laporta:1996mq}
  S.~Laporta and E.~Remiddi,
  %``The analytical value of the electron (g-2) at order alpha~3 in QED,''
  Phys.\ Lett.\  B {\bf 379}, 283 (1996)
  [arXiv:hep-ph/9602417].
  %%CITATION = PHLTA,B379,283;%%

\bibitem{Broadhurst:1996yc}
  D.~J.~Broadhurst and A.~V.~Kotikov,
  %``Compact analytical form for non-zeta terms in critical exponents at  order
  %1/N**3,''
  Phys.\ Lett.\  B {\bf 441}, 345 (1998)
  [arXiv:hep-th/9612013].
  %%CITATION = PHLTA,B441,345;%%

\bibitem{Kniehl:2006bf}
  B.~A.~Kniehl and A.~V.~Kotikov,
  %``Heavy-quark QCD vacuum polarisation function: Analytical results at  four
  %loops,''
  Phys.\ Lett.\  B {\bf 642}, 68 (2006)
  [arXiv:hep-ph/0607201].

\bibitem{Kniehl:2006bg}
  B.~A.~Kniehl, A.~V.~Kotikov, A.~I.~Onishchenko, and O.~L.~Veretin,
  %``Strong-coupling constant with flavor thresholds at five loops in the
  %MS-bar scheme,''
  Phys.\ Rev.\ Lett.\  {\bf 97}, 042001 (2006)
  [arXiv:hep-ph/0607202].
  %%CITATION = PRLTA,97,042001;%%

\bibitem{PSLQ}
  H.~R.~P.~Ferguson and D.~H.~Bailey, RNR Technical Report No.\ RNR-91-032;
  H.~R.~P.~Ferguson, D.~H.~Bailey and S.~Arno, NASA Technical Report No.\ 
  NAS-96-005.

\bibitem{GraRy}
I.S. Gradshteyn and I.M. Ryzhik, {\it Table of Integrals, Series and Products}
(Academic Press, New York, 1980).

\bibitem{Kazakov:1987jk}
  D.~I.~Kazakov and A.~V.~Kotikov,
  %``TOTAL ALPHA-s CORRECTION TO DEEP INELASTIC SCATTERING CROSS-SECTION RATIO,
  %R = sigma-L / sigma-t IN QCD. CALCULATION OF LONGITUDINAL STRUCTURE
  %FUNCTION,''
  Nucl.\ Phys.\  {\bf B307}, 721 (1988);
{\bf B345}, 299(E) (1990);
  %%CITATION = NUPHA,B307,721;%%
%\bibitem{Kazakov:1987jm}
%  D.~I.~Kazakov and A.~V.~Kotikov,
  %``TOTAL ALPHA-s CORRECTION TO DEEP INELASTIC SCATTERING CROSS-SECTION RATIO,
  %R = sigma-L / sigma-t IN QCD. GAUGE INVARIANCE AND COMPARISON WITH
  %EXPERIMENT,''
%  Nucl.\ Phys.\  B {\bf 345}, 299 (1990);
  %%CITATION = NUPHA,B345,299;%%
%\bibitem{Kazakov:1986mu}
%  D.~I.~Kazakov and A.~V.~Kotikov,
  %``THE METHOD OF UNIQUENESS: MULTILOOP CALCULATIONS IN QCD,''
  Theor.\ Math.\ Phys.\  {\bf 73}, 1264 (1988)
  [Teor.\ Mat.\ Fiz.\  {\bf 73}, 348 (1987)];
  %%CITATION = TMFZA,73,348;%%
%\bibitem{Kotikov:1987mw}
  A.~V.~Kotikov,
  %``THE CALCULATION OF MOMENTS OF STRUCTURE FUNCTION OF DEEP INELASTIC
  %SCATTERING IN QCD,''
  Theor.\ Math.\ Phys.\  {\bf 78}, 134 (1989)
  [Teor.\ Mat.\ Fiz.\  {\bf 78}, 187 (1989)].
  %%CITATION = TMFZA,78,187;%%

\bibitem{Kalmykov:2008ge}
  M.~Yu.~Kalmykov and B.~A.~Kniehl,
  %``Towards all-order Laurent expansion of generalized hypergeometric functions
  %around rational values of parameters,''
  Nucl.\ Phys.\ {\bf B809}, 365 (2009)
  [arXiv:0807.0567 [hep-th]].
  %%CITATION = NUPHA,B809,365;%%

\end{thebibliography}
\end{document}